\documentstyle[epsfig]{mn}
\input{psfig}
\begin{document}
\def\ltsima{$\; \buildrel < \over \sim \;$}
\def\simlt{\lower.5ex\hbox{\ltsima}}
\def\gtsima{$\; \buildrel > \over \sim \;$}
\def\simgt{\lower.5ex\hbox{\gtsima}}

\title[The X--ray spectrum and variability  
of the Seyfert 2 Galaxy NGC 7172]
{The X--ray spectrum and variability  
of the Seyfert 2 Galaxy NGC 7172}

\author[M. Guainazzi et al.]
{M. Guainazzi$^{1,2,*}$, G. Matt$^3$, L. A. Antonelli$^{1}$,  
F. Fiore$^{1,4}$, L. Piro$^5$,
and S. Ueno$^6$ \\ ~ \\
$^1$ SAX/SDC Nuova Telespazio,
Via Corcolle 19,  I--00131 Roma, Italy \\
$^2$ Unit\`a GIFCO/CNR, Via Archirafi 36, I-90123 Palermo, Italy \\
$^3$ Dipartimento di Fisica, Universit\`a degli Studi ``Roma Tre", 
Via della Vasca Navale 84, I--00146 Roma, Italy \\
$^4$  Osservatorio Astronomico di Roma, Via dell'Osservatorio,
I-00044 Monteporzio-Catone, Italy \\
$^5$  Istituto di Astrofisica Spaziale, CNR,
Via Fosso del Cavaliere, I-00133 Roma, Italy \\
$^6$ X--ray Astronomy Group, Department of Physics \& Astronomy, University of
Leicester, University Road, LE1 7RH, Leicester, United Kingdom \\
$^*$current address: Astrophysics Division, Space Science Department of
ESA, ESTEC, Postbus 299, 2200 AG Norordwijk, The Netherlands\\
}

\maketitle
\begin{abstract}
We present evidence of flux variability, on both short (hours) and 
long (months) time scales of the Seyfert 2 galaxy
NGC 7172. These results are based on the ASCA observation of NGC 7172
performed on May 1996. The source was detected at a rather low flux level, about
3 times fainter than its usual state (including one year before, when it
was also observed by ASCA).

The source also varied by about 30 percent during the observation, confirming
the presence of a type 1 nucleus in its center. 
However, its spectrum appears to be
flatter than the typical Seyfert 1 spectrum (in agreement with findings on
other Seyfert 2's), posing problems for 
the unification model unless complex absorption is invoked. 

\end{abstract}

\section{Introduction}

NGC 7172 is a S0-Sa (Anupama et al. 1995)
Seyfert 2 galaxy (Sharples et al. 1984) seen almost edge--on; it belongs
to a compact group, HCG 90. 

It is one of the brightest Seyfert 2's in X--rays, being one of the 
Piccinotti sources, and 
it has been observed by most 
satellites (see Polletta et al. 1996a for a recent catalog of
X--ray measurements of Seyfert 2 galaxies), 
namely Ariel V/SSI (McHardy et al. 1981), HEAO1/A1 (Wood et 
al. 1984) and A2 (Marshall et al. 1979; Piccinotti et al. 1982), EXOSAT
(Turner \& Pounds 1989), GINGA (Warwick et al. 1993; Nandra \& Pounds 1994;
Smith \& Done 1996), ROSAT/PSPC (Polletta et al. 1996a). The 2--10 keV
flux has been fairly constant during the 1977-1989 period, at a level of about
3--4$\times$10$^{-11}$ erg cm$^{-2}$ s$^{-1}$,
corresponding to an unabsorbed luminosity of about 1.5--2$\times$10$^{43}$
erg s$^{-2}$. No compelling
evidence for an iron K$\alpha$ line has been observed yet, the best measurement
being a 40$\pm$40 eV equivalent width value obtained with GINGA 
(Nandra \& Pounds 1994). The same observation provided, 
however, evidence for a Compton reflection 
continuum in the form of the hardening of the spectrum above 10 keV
(the power-law photon spectral index was measured to be
1.80$\pm$0.03 in the 2--18 keV band, and 1.52$\pm$0.11 in the 
10--18 keV band). As ubiquitous in X--ray detected
Seyfert 2's, the absorption is largely in excess of the 
Galactic one, being about 10$^{23}$ cm$^{-2}$ according to GINGA.

Its soft (0.1--2.4 keV) X--ray luminosity, 
as measured by ROSAT/PSPC (Polletta et al., 1996),
is 2.6$\times$10$^{40}$ erg s$^{-1}$. This emission could in principle 
originate 
in an extended region, like for instance NGC~4388 (Matt et al. 1994) which
has a similar soft X--ray 
luminosity, or by scattering of the nuclear radiation off
warm matter. However, as the galaxy belong to an X--ray emitting group
(Ponman et al. 1996), it is possible that this emission is actually due
to the intergalactic gas. We will return later to this point. 

ASCA (Tanaka et al. 1994) observed this source twice: on 13 May 1995 
(Ryde et al., 1996, 1997) 
and on 17 May 1996. The results from the latter observation, as well as
re-analysis of the ROSAT HRI data, are presented here (preliminary 
results can be found in Matt et al., 1996)

\section{Data Analysis and Results}

Effective observing time 
of May 1996 observation
was about 20 ksec. Data reduction, selection and analysis
were performed using the FTOOLS 3.6 and XANADU 9.0 software
packages, and adopting standard
selection criteria (the results do not differ appreciably from those 
obtained using the previous (3.5) 
FTOOLS version, and reported in  Matt et al., 1996).
In particular, we used only data obtained 
out of the South Atlantic Anomalies, 
with an elevation angle greater than 10 degrees towards the dark
limb of the Earth and 20 degrees towards the bright limb. The SIS data
were operating in BRIGHT2 mode;  Poissonian--based automatic cleaning of
hot and flickering chip pixels was performed. 
Background subtraction for the 2--10 keV analysis has been performed
using the blank sky (BS) archive observations, after checking that no 
differences arise, in this band, if background is taken in an empty 
(of point sources) zone in the observation field of view (see Sec.2.3 for the
background subtraction below 2 keV). 

Despite NGC~7172 is in a group, no other point--like sources were evident
in the image.

\subsection{2--10 keV spectrum}

The source was, in our observation, much fainter than usual; its observed
2--10 keV flux was in fact $\sim$1.3$\times$10$^{-11}$ erg cm$^{-2}$ s$^{-1}$,
i.e. 3--4 times fainter than in previous observations, including the ASCA one
performed exactly one year before (Ryde et al. 1997). Therefore,
after having checked that all the spectra were consistent, we
fitted all four instruments (namely 
the two CCD's, SIS0 and SIS1, and the two GSPC's,
GIS2 and GIS3) together
in order to maximize the signal to noise ratio. The overall normalization
of the two GIS have been left free to vary with respect to that of the two SIS
to allow for possible miscalibrations between the two detectors.
The fits discussed in this section have been performed using only data
above 2 keV, when  the nuclear radiation is directly visible. Soft
X--rays are discussed below. 

\begin{figure}
\epsfig{figure=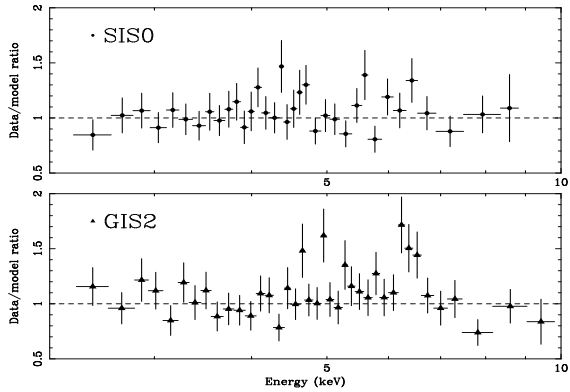,height=8.5cm,width=5.5cm,angle=-90}
\caption{Data/model ratio when a simple absorbed power-law
model is applied to the SIS0 ({\it upper panel}) and GIS2 ({\it lower
panel}) NGC7172 spectra}
\label{fig6}
\end{figure}

\begin{figure}
\epsfig{figure=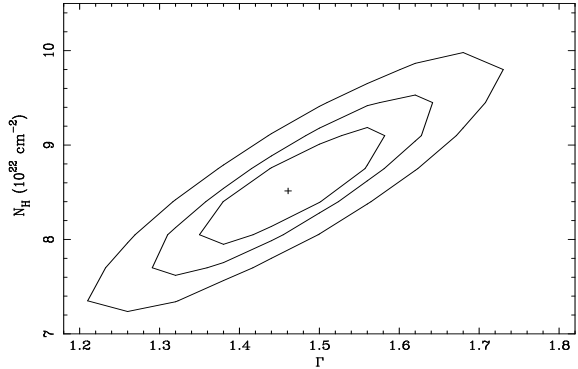,height=8.5cm,width=5.5cm,angle=-90}
\caption{$\Gamma$ vs. $N_H$ contour plot when a simple absorbed power-law
model is applied simultaneously  to the NGC7172 ASCA observation spectra of the four
detectors in the 2--10~keV range}
\label{fig7}
\end{figure}

The results of our 2--10 keV spectral analysis are summarized in Table 1.
\begin{table*}
\centering
\caption{
Best fit parameters for the May 1996 observation. Errors correspond to 
$\Delta\chi^2$=2.7. The iron line energy is given in the source rest frame.}
\label{all}
\vspace{0.05in}
\begin{tabular}{lcccccccc}
\hline
\# & N$_{\rm H}$ (10$^{22}$ cm$^{-2}$) & $\Gamma$ & E$_{\rm line}$ (keV) & 
$\sigma_{\rm line}$ (keV) & E.W. (eV) & N$_{\rm H, pc}$ (10$^{22}$ cm$^{-2}$) 
&  F$_{\rm c}$ & $\chi^2_r$/d.o.f. \cr
\noalign {\hrule}
1 & 8.52$^{+0.83}_{-0.76}$ & 1.46$^{+0.15}_{-0.14}$ & -- & -- & -- & -- &
-- & 0.98/448 \cr
2 &10.3 & 1.74$^{+0.08}_{-0.07}$ & -- & -- & -- & -- & -- & 1.00/449 \cr
3 & 8.61$^{+0.79}_{-0.33}$ & 1.52$^{+0.14}_{-0.15}$ & 6.39$^{+0.10}_{-0.10}$
   & 0.01 & 121$^{+50}_{-60}$ & -- &  -- & 0.95/446 \cr
4 & 10.3 & 1.79$^{+0.09}_{-0.10}$ & 6.40$^{+0.12}_{-0.12}$ & 0.01 & 
  160$^{+31}_{-110}$ & --  & -- & 0.98/447 \cr
5 & 8.5$^{+0.8}_{-2.1}$  & 1.61$^{+0.20}_{-0.20}$ & 5.99$^{+0.33}_{-0.42}$ & 
  0.85$^{+4.69}_{-0.46}$  & 480$^{+970}_{-280}$   & -- & -- &
0.95/445 \cr
6 & 9.03$^{+1.47}_{-3.34}$ &  2.20$^{+0.38}_{-0.30}$ & -- & -- & -- & 
26.6$^{+18.3}_{-16.8}$ & 0.58$^{+0.11}_{-0.16}$ & 0.94/446 \cr
7 & 6.9$^{+2.8}_{-6.9}$ &  1.93$^{+0.18}_{-0.32}$ & 6.40$^{+0.11}_{-0.10}$
 & 0.01 & 99$^{+49}_{-64}$ & 
11.5$^{+21.7}_{-5.5}$ & 0.62$^{+0.36}_{-0.26}$ & 0.93/444 \cr
\hline
\end{tabular}
\end{table*}
We firstly 
fitted the spectrum with a simple absorbed power law (model 1 in the Table),
which provides a satisfactory result, as far as the $\chi^2$ is concerned.
In Figure~\ref{fig6} the ratio data/model is shown for the SIS0 and GIS2
data.
The spectrum is highly absorbed, a fact already known for this source and
very common for Seyfert 2 galaxies. 
The spectral index appears to be rather flat 
($\Gamma$=1.46$^{+0.15}_{-0.14}$;
hereinafter all errors correspond to 90$\%$ confidence level for one
interesting parameter, unless otherwise specified),
significantly flatter than the mean value
for Seyfert galaxies
(1.73, as  obtained by 
Nandra \& Pounds 1994 for their sample of Seyfert galaxies, when fitting the
spectra with a simple power law), 
and also than the value obtained for this source by
EXOSAT and GINGA ($\sim$1.8, 
Turner \& Pounds 1989; Nandra \& Pounds 1994). It is
however consistent with that found by Ryde et al. (1997) in the 1995 ASCA
observation ($\Gamma$=1.47$\pm$0.15), i.e. when the source was about four times
brighter. If we fix 
the column density of the absorbing material to the GINGA value 
(i.e. 10.3$\times$10$^{22}$ cm$^{-2}$, 
Nandra \& Pounds 1994), we obtain a spectral index
consistent with the GINGA one 
(see model 2 in the table); the fit, even if still statistically acceptable,
is worse with respect to the previous one at more than 99.9 percent confidence 
level, on the ground of the F--test. This can also be seen from Figure~\ref{fig7}
where the $\Gamma$-N$_{\rm H}$ contour plot is shown: the 1.8 value is 
excluded at the 99.7 percent confidence level. Moreover, fitting 
the four instruments separately with the column density left free, flat
spectra are always preferred by the fitting procedure. So, we conclude that
a flat spectrum is favored.

The result is rather stable when fits are performed on narrower energy
ranges. Since a $8.5 \times 10^{22} \ cm^{-2}$ absorbing column
density has a low-energy e-folding $E_C \simeq 3 \ keV$, best-fit
$N_H$ assumes high and poorly constrained values
as the lowest limit of the range where the fit is performed approaches
or overheads $E_C$
({\it e.g.}: $N_H = 18^{+7}_{-3} \times 10^{22}$ if the fit is
performed in the 4--10~keV range. This measure is not consistent with
anyone obtained y X--ray experiments so far). If we assume the measure of
$N_H$ in the 2--10~keV to be correct, the spectral index
becomes indeed steeper with a narrower band, but still
consistent with Ginga results and flat
($\Gamma_{3--10 \ keV} = 1.52 \pm 0.08$,
$\Gamma_{4-10 \ keV} = 1.57 \pm 0.12$).

To test whether this flatness could be an artifact due to 
complex absorption, we have fitted the spectrum with a
dual absorber. One absorber is assumed to cover the whole source, 
while the other one only partially. The results are reported
in Table 1 (model 6 and 7). The improvement in the fit is significant but 
not dramatic, and the parameters are not very well constrained
(especially the column densities), owing
to the modest signal to noise of the present observation;
the spectral index turns out to be steep, even too much (when compared
with the mean value of Nandra \& Pounds 1994),  especially if the iron line
is not included in the fit procedure. 
To settle the issue of the real nature of the flat spectrum, 
broad band simultaneous observations, like those performed by Beppo--SAX 
and RXTE, seem mandatory. 

\subsection{Fluorescent iron line and reflection component}

\begin{figure}
\epsfig{figure=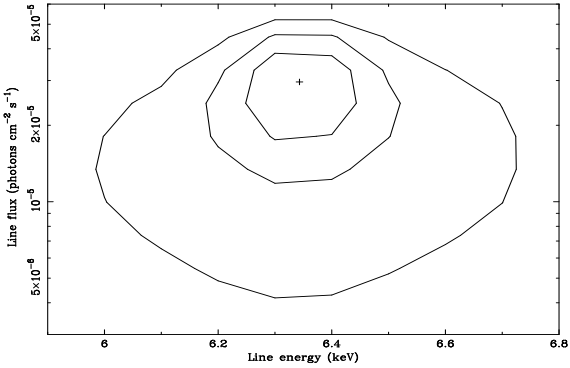,height=8.5cm,width=5.5cm,angle=-90}
\caption{Iron line flux versus energy confidence contours for an absorbed
power-law + narrow Gaussian line model. Confidence levels are
68\%, 90\% and 99\% for 2 interesting parameters}
\label{fig8}
\end{figure}

Even if the simple power law provides a satisfactory fit to the data,
an excess emission between 6 and 7 keV is apparent (Fig.~1) in the GIS data
(i.e. the most sensitive instruments aboard ASCA at those energies).
We then tried to include in the fit also an iron fluorescent
emission line, which is a common feature in Seyfert galaxies, and is sometimes
very strong
in Seyfert 2's. We added a narrow Gaussian feature to both 
the single and dual absorbed power law models. 
The results are shown in Table 1 (models 3, 4 and 7). The 
improvement in the $\chi^2$ when the line is added 
ranges from $\Delta \chi^2 = 11$ to $\Delta \chi^2 = 26$.
The contour plots of the line flux against
both the line energy (see Figure~\ref{fig8}) and $\Gamma$
suggest that a line is
actually present at a confidence level larger than 99.9\% for 2 interesting
parameters.

\begin{figure}
\epsfig{figure=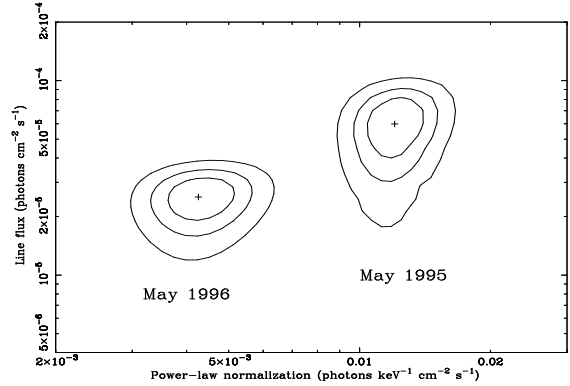,height=8.5cm,width=5.5cm,angle=-90}
\caption{Iron line flux vs power-law normalization
contours for the May 1995 and May 1996
observations. Confidence levels are 1 $\sigma$, 68\% and 90\% for
two interesting parameters ($\Delta \chi^2 = 4.61$)}
\label{fig9}
\end{figure}

The best fit values for the line equivalent width (EW)
ranges from 99 to 164 eV, depending on 
the modeling of the absorption. The value obtained with the simple absorber
and with the column density left free to vary 
is 121 eV, which intriguingly is exactly three times
higher than the best fit GINGA value (40 eV, Nandra \& Pounds 1994), obtained
when the continuum was three times higher. On the other hand, a re--analysis
of the May 1995 ASCA data (which are now in the public archive) revealed that
the equivalent width is instead approximately the same for the two observations,
suggesting that the line flux follows the continuum and therefore 
a line origin close to the continuum source
(see Figure~\ref{fig9}). However, the line parameters
are so poorly constrained that no firm conclusions can actually be reached. 

Broad iron lines from the innermost region of the accretion disc have also
been observed in many Seyfert 1's (Tanaka et al. 1995; Nandra et al. 1997).
Leaving the line width free to vary (model 5), 
the fit suggests a very broad, redshifted and strong
line, which would be consistent with what is expected from an 
accretion disc rotating around a Kerr black hole (see Martocchia \& Matt
1996 and references therein). However, there is not a significant
improvement in the statistical quality of the fit; any further study
on a broad disc component must be deferred to higher quality data.

We searched also for the presence of a Compton reflection continuum
(Lightman \& White 1988; George \& Fabian 1991;
Matt, Perola \& Piro 1991), as such a component is expected to go along with
any line emission (if originating from reflection off optically thick matter)
and could at least partly explain the flatness of the spectrum.
Unfortunately, the
limited bandwidth of ASCA is not very well suited to this task, 
and moreover the rather poor signal to noise of the
present observation does not help in studying this component. 
We nevertheless tried to add a reflection component
to model 3. No significant improvement
in the goodness of the fit is found, and the best fit value for the 
relative normalization of the Compton reflection component, 
{\it R}, is very high, 19$^{+38}_{-12}$, a rather
unrealistic value (note that a huge iron line should be present in this 
case). We therefore conclude that any reflection component is basically
unconstrained; again broadband missions measurements will
be valuable in this respect.

\subsection{Soft X--rays}

\begin{figure}
\epsfig{figure=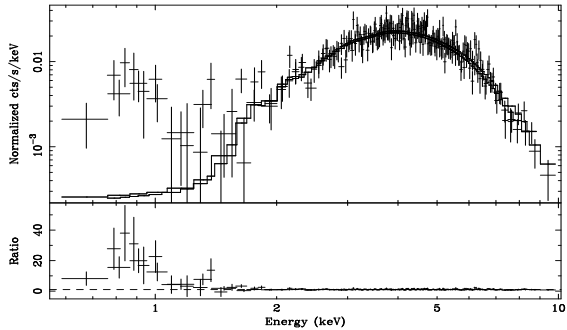,height=8.5cm,width=5.5cm,angle=-90}
\caption{SIS spectra ({\it upper panel} and residuals ({\it lower panel})
when a simple absorbed power-law model is applied in the 0.57--10~keV energy
range. Note the prominent soft excess}
\label{fig10}
\end{figure}

The ASCA spectrum below $\sim$2 keV depends dramatically on the 
choice of the background to be subtracted. If background spectra
are extracted from an outer region of the same field of view as the
source, the absorbed
power law provides a good fit to the whole SIS band; on the
contrary, if the 
BS background is chosen, a clear soft X--ray excess is present (see
Figure~\ref{fig10}).
No obvious instrumental explanations do exist for this effect.
Recalling that NGC~7172
belongs to a compact group (Hickson 90: H90), the best explanation seems that
the soft excess have to be attributed to extended emission from the group
rather than to the active nucleus; subtracting a local background, this excess
therefore disappears. Note that H90 belongs to the Ponman et al. (1996)
ROSAT sample; they fitted the spectrum with a 
thermal plasma model which gives a temperature of 0.68$\pm$0.12~keV,
a metal abundance of  0.20$\pm$0.16 in solar units, and a total X--ray 
luminosity within a radius of 200 kpc (i.e. 13' at the distance
of H90, which is  52.8 Mpc assuming H$_0$=50 km s$^{-1}$ Mpc$^{-1}$)
of 3$\times$10$^{41}$ erg s$^{-1}$.  We have then analyzed the soft X--rays
properties of the group using ASCA SIS's, and re-analyzed higher
spatial resolution ROSAT HRI archive observations. 

\subsubsection{ASCA}

To derive the spectral properties of the intragroup plasma emission, and to 
study any possible contribution from NGC~7172 itself, we analyzed both the SIS
4' regions including the active galaxy and the 2' regions originally
selected to estimate the local background. Subtracting the blank sky background
from the local background, a clear excess below 1 keV is present, which can
be well fitted with a Raymond--Smith plasma model. The metal abundance and
the column density  have been fixed to 0.2 and to the Galactic value
(1.5$\times10^{20}$ cm$^{-2}$) respectively. The best fit temperature is about
0.70, consistent with the ROSAT value (Ponman et al. 1996). We have
also tried to leave free the abundance, but the parameter was
basically unconstrained. Similar results are obtained by fitting simultaneously
a Raymond--Smith and an absorbed power law over the whole SIS band for the
spectrum extracted from the region containing NGC~7172.

\begin{figure}
\epsfig{figure=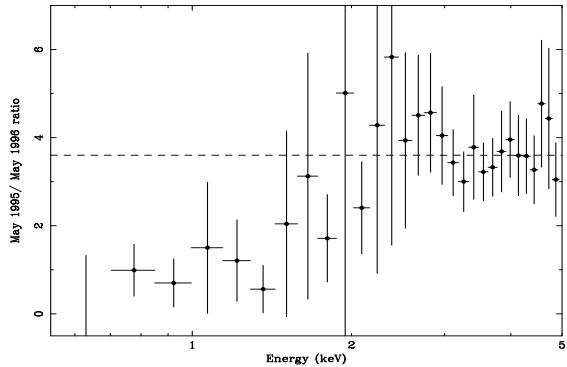,height=8.5cm,width=5.5cm,angle=-90}
\caption{Ratio of the May 1995 and May 1996 observations
SIS1 spectra in the 0.54--5~keV energy band. The
dashed lines marks the 2--10~keV flux ratio. Soft X--ray emission below 1.5 keV
does not show any significant variability}
\label{fig11}
\end{figure}

The soft X--ray emission seems not to be significantly variable, despite more
a factor of 3 flux changes in the 2--10~keV band. In Figure~\ref{fig11} the
ratio of the SIS1 spectra from the May 1995 and May 1996 ASCA observations is
shown. The ratio is consistent with 1 for $E \simlt 1.5 \ keV$, while the
flux ratio for $E > 2 \ keV$ is about 3.6. 

\subsubsection{ROSAT--HRI}

\begin{figure}
\epsfig{figure=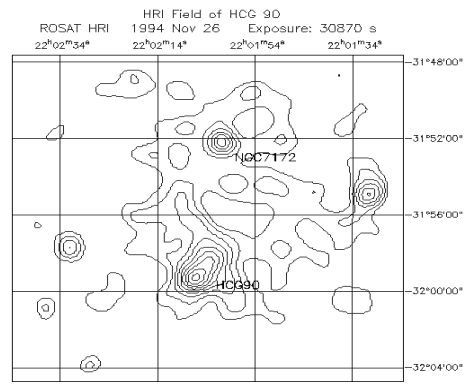,height=8.5cm,width=5.5cm,angle=-90}
\caption{Intensity contour of the central region
in the HRI observation of the H90 group. The
X--ray center of the group and the location of NGC7172 are indicated.}
\label{fig1}
\end{figure}

The HCG90 group was observed by the HRI detector on board ROSAT from 26 Nov 
1994 up to 29 Nov 1994 for a total observation time of 30870 s.
In Figure~\ref{fig1} an intensity contour map of a subset of the HRI field is
shown.
In total eleven sources are detected in the
field. Six of these are present in White, Giommi and Angelini
(WGA, 1994) 
catalog and five are serendipitous sources. 
The strongest source, located at about 11 arcmin South from the center 
(R.A.:$22^{h}$ $02^{m}$ $32.4^{s}$ Dec.:$-32^{o}$  $08'$  $1".7$), is a F6/F7V
spectral type star identified as SAO 213495 and observed to emit X-rays for 
the first time by EXOSAT (Tagliaferri et al., 1994).
The second brightest source is a serendipitous
located near the west border
(R.A.: $22^{h}$ $01^{m}$ $31.5^{s}$, Dec.: $-31^{o}$  $54'$  
$54".1$). The only known source in its neighborhood is the group of galaxies
EDCC76 (Lumsden et al., 1992) which lies at about 1 arcmin off the ROSAT
position.

A faint source, 
$[(7.0\pm2.0)\times10^{-4} \ cts \ s^{-1}]$, is located at the center of the group; this is present in WGA catalog and in IRAS Pointing Sources Catalog too.
NGC7172 lies at about 6.5 arcmin off the X-ray center of the group
and is distant at least 5 arcmin from any other detected source in the field.

\begin{figure}
\epsfig{figure=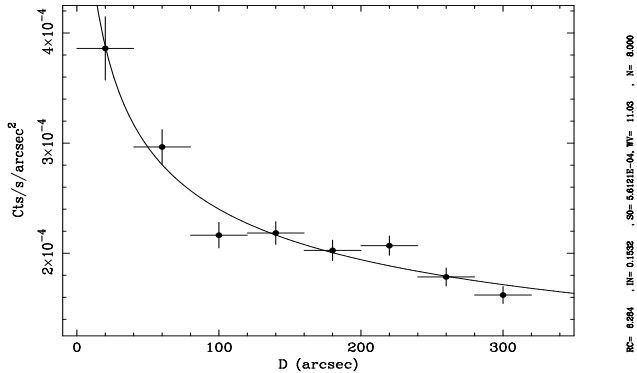,height=8.5cm,width=5.5cm,angle=-90}
\caption{Radial surface brightness profile for H90. The solid line represents
the best fit King's model}
\label{fig5}
\end{figure}

\begin{figure}
\epsfig{figure=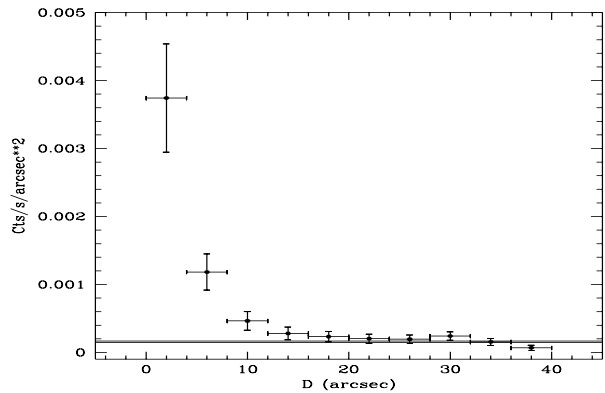,height=5.5cm,width=8.5cm}
\caption{Surface brightness radial profile for NGC7172. Solid lines
represent the 90\% confidence limits for the contribution of H90
assuming the King's profile best fit parameters given in text. No
significant evidence of intrinsic broadening above the HRI point
spread function is present}
\label{fig2}
\end{figure}

We computed the radial surface brightness profile of the group,
in order to estimate the contribution to NGC7172 soft X-ray emission.
All the sources whose count rate had a chance
probability $<10^{-4}$ to be due to a fluctuation of the local background
were removed from the field.
The radial profile has been calculated from equally spaced 40'' wide ({\it i.e.}
$R_{out}-R_{in}$) annulus. The instrumental background count rate
has been estimated as the arithmetic mean of the
brightness obtained from three areas of the field of view which were free
from any apparent contaminating source, provided they yielded consistent
results within the statistical uncertainties. In Figure~\ref{fig5} the
brightness radial profile for H90 is shown. It can be well fit with
a King's profile $B = N (1+(R/R_c)^2]^{-\alpha}$
(King 1962). The best-fit
parameters are $N=(5.61\pm0.16) \times 10^{-4} \ s^{-1} \ arcsec^{-1}$,
$R_c = 6.3 \pm 0.6 \ arcsec$ and $\alpha \simeq 0.153$ (no meaningful
constraint could be derived for the last parameter).
In Figure~\ref{fig2} the surface
brightness radial profile centered on the apparent centroid of NGC7172
is plotted; on the same plot is superimposed the expected contribution
of the underlying cluster according to the best-fit King's model
at the observed $\sim$ 6.5 arcmin distance. The NGC7172 brightness is
consistent with 0 from $\sim$ 12 arcsec outwards and decreases by a factor
$\simeq$ 10 in the inner 12 arcsec. Such a finding is consistent with
the expected count decrement due to the instrumental point spread function.

NGC7172 is clearly visible in the HRI image. The net
count rate when the
underlying group contribution is
subtracted is $2.0 \pm 0.5 \times 10^{-3} \ s^{-1}$
(only data points between the channel
2 and 9 included have been used). Using tool {\sc PIMMS} we have estimated
the expected count rate for a simple power-law of spectral index 1.8,
2--10~keV flux $\sim 6 \times 10^{-11} \ erg \ s^{-1} \ cm^{-2}$
and absorbed by a $8 \times 10^{22} \ cm^{-2}$ column density as $\simeq
1.8 \times 10^{-3} \ s^{-1}$ (the best-fit parameters are taken from
Ginga results and are supposed to be representative of the spectral state
of the source when the HRI observation was performed). HRI data
of NGC7172 do not
therefore
require any spatially unresolved component but the high-energy power-law.

The above results lead us to conclude
that the soft excess in the ASCA spectrum of NGC7172 can
be accounted for by the intergalactic gas of the group and
neither an extended component
related to the galaxy, nor an emission from the nuclear environment
different from the nuclear high-energy power-law
is needed.

\section{Variability}

\subsection{Short term variability}

\begin{figure}
\epsfig{figure=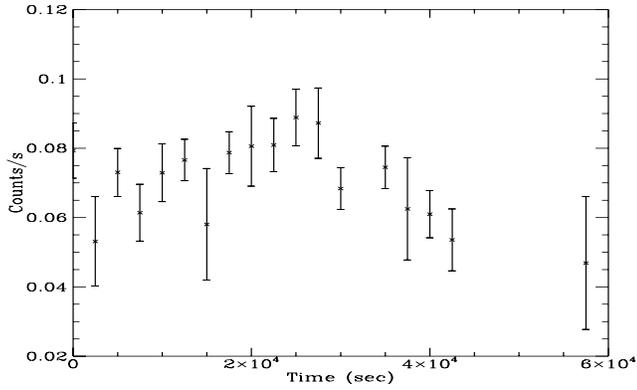,height=5.5cm,width=8.5cm}
\caption{SIS1 NGC7172 light curve (binning time $\Delta t = 2500 \ s$}
\label{fig4}
\end{figure}

During the May 1996 observation the source exhibited a flux variability
of about 30 percent on time scales of hours. The light curve, binned 
at 2500 s (i.e. about half an orbit) of the
SIS1 is presented in Figure~\ref{fig4}. The light curves of the
other instruments are very similar, while the background stays constant.
This variability implies an 
X--rays emitting region smaller than a few$\times$10$^{14}$ cm, or
$\sim$100$r_{\rm g} M_7$, where $r_{\rm g}=GM/c^2$ and 
$M_7$ is the black hole mass in units of 10$^7$ solar masses. 
In other words, in hard X--rays we are looking at the innermost hundred 
gravitational radii or so, i.e. at 
the very nucleus of the source which, as far as time variability and
luminosity are concerned, appears very much like a Seyfert 1 nucleus, 
in agreement with unification schemes.

\subsection{Long term variability}

\begin{figure}
\epsfig{figure=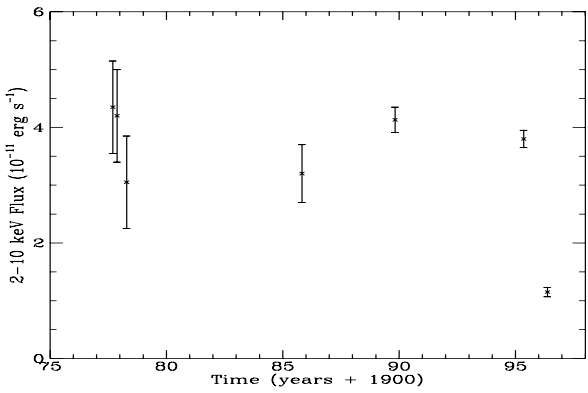,height=5.5cm,width=8.5cm}
\caption{Historical light curve of NGC7172, including HEAO-1, Ariel-V,
Ginga, EXOSAT and
the two ASCA observations}
\label{fig3}
\end{figure}

In May 1996 ASCA caught the source in an unprecedented low flux state.
In fact, since the first X--ray observations of NGC 7172, 
performed in the seventies by
HEAO-1 and Ariel V (see Polletta et al. 1996 and references therein), and
till at least May 1995, the source was always observed at a flux 
level between $\sim$3 and $\sim$5$\times$10$^{-11}$ erg cm$^{-2}$ s$^{-1}$, 
while in May 1996 it was at a flux of 
about 1.3($\pm$0.1)$\times$10$^{-11}$ erg cm$^{-2}$ s$^{-1}$. 
The flux history of NGC 7172 is summarized in Figure~\ref{fig3}.
Line emission is not very well constrained but, adopting
the best fit values, it 
appears to follow the long term flux variability, at least if 
the two ASCA observations,  performed about 1 year apart, are compared.
This would imply that the line emitting region is closer to the continuum
emitting region than 1 light--year,
corresponding to 0.3~pc.
This measure is still consistent with a torus size. In the case of NGC1068,
for instance, 0.3--0.5~pc for the inner edge size of the molecular torus were
suggested (Greenhill et al. 1996, Gallimore et al. 1997).
However, the EW of an iron line as seen in transmission through a
Compton-thin medium is $\sim 30 f_c$, where $f_c$ is the
covering fraction of the torus
(Ghisellini, Haardt \& Matt, 1994). It is therefore unlikely that
all the observed line is accounted by the torus this way, although
it cannot be {\it a priori} ruled out that a contribution from it
comes. 

Alternatively, the line could be generated in an accretion
disk and therefore much closer to the central black hole. In this
case the observed EW is consistent with the observed in
the Nandra et al. (1997) sample of Seyfert 1 galaxies.
Were this true,, the line profile should suffer broadening by relativistic
and kinematics effects, as commonly observed in Seyfert 1's.
As discussed in Sec.2.2, however, present data do not permit
to constrain the line width. 
As the line peaks at 6.4~keV, this scenario implies a face-on disk
(Matt, Perola \& Piro, 1991), whereas the unification picture would
predict an disk seen edge-on in Seyfert 2s. VLA observations did not
detect any clear extended radio source within NGC7172
(Condon {\it et al.} 1996) and then do not allow to tell whether
an highly inclined disk indeed exists. Moreover, the average iron line
profile in a sample of Seyfert2s observed by ASCA (Turner {\it
et al.}, 1997) shows a very similar red wing as in Seyfert 1s, thus
suggesting a similar projected geometry of the emitting regions.
This is of course not in line with the 0-th order unification model.
 
\section{Discussion and conclusions}

We have analyzed the ASCA observation of NGC~7172 performed in May 1996.
The main results can be summarized as follows:

\noindent
a) The source was 3--4 times fainter than usual; in particular it was
4 times fainter than in May 1995, when another ASCA
observation was performed (Ryde et al. 1996, 1997).

\noindent
b) During the observation, the source varied by about 30 percent on time
scales of hours. Such a short term variability supports the hypothesis 
of the presence of a type 1 nucleus. 

\noindent
c) The spectrum appears to be rather flat, i.e. $\Gamma\sim$1.5, and highly 
absorbed ($N_{\rm H}\sim8\times10^{22}$ cm$^{-2}$). 
Such a spectral index is significantly 
flatter than that detected by GINGA and EXOSAT, but consistent
with that observed in the previous ASCA observation. A steeper spectral
index with either a higher value of the column density (i.e. fixed to
that observed by GINGA) or a huge amount
of Compton reflection cannot be ruled out. However,
the first case is disfavored on statistical ground, while in the second case
a very large iron line, much stronger than observed, would be expected. 
A third possibility is that the flat spectrum is actually due to complex
absorption. A dual absorber is accepted but not strongly required by the data. 

It is worth noticing that there are other Seyfert 2 galaxies 
with a flat spectrum
(see Cappi et al. 1995; Smith \& Done 1996). If intrinsic, this flatness
would be 
a problem for the unification model (see Antonucci 1993 for a review), 
at least in its simplest version (even if the
short term variability of this same source suggests, on the contrary, that 
we are observing in hard X--rays a type 1 nucleus). Whether the flatness
of the spectrum is due to an intrinsic difference between 
Seyfert 1's and 2's, or
reflects an angular dependence of the spectral shape (with high inclination
systems, as Seyfert 2's are expected to be in the unification scenario, 
being flatter) or finally is
an artifact of the absorption being actually complex, cannot be said
with the present data alone. Observations of Seyfert 2 galaxies at higher 
energies seems mandatory in this respect.

\noindent
d) An iron fluorescent line with EW$\sim$120$\pm$60 eV is also present. The
EW is consistent with that of the previous ASCA observation, suggesting that
the line flux follows the continuum flux. 

\noindent
e) A soft excess at energies $E \simlt 2 \ keV$ can be attributed to the
compact group H90, within which NGC7172 lies, and no extended
emission connected to the galaxy nearby surrounding environment
or nuclear emission above the high-energy power-law is
required.

The present ASCA observation of NGC~7172 has raised some interesting 
questions about the emission of this source and of Seyfert 2 galaxies at
large. Many of these
questions require higher energy measurements and therefore 
may will be addressed by forthcoming RXTE and BeppoSAX observations.

\section*{Acknowledgments}

The authors acknowledge useful discussions with F.Ryde and valuable
suggestions from an anonymous referee.

\end{document}